*Article*

# Comparing Multi-Walled Carbon Nanotubes and Halloysite Nanotubes as Reinforcements in EVA Nanocomposites

**Agata Zubkiewicz [1],\*, Anna Szymczyk [1], Piotr Franciszczak [2], Agnieszka Kochmanska [2], Izabela Janowska [3] and Sandra Paszkiewicz [2,\*]**

[1] Department of Technical Physics, West Pomeranian University of Technoclogy, 70311 Szczecin, Poland; aszymczyk@zut.edu.pl
[2] Department of Materials Technology, West Pomeranian University of Technology, 70310 Szczecin, Poland; piotr.franciszczak@zut.edu.pl (P.F.); akochmanska@zut.edu.pl (A.K.)
[3] Institut de Chimie et Procédés pour l'Energie l'Environnement et la Santé (ICPEES), University of Strasbourg 67087, France; janowskai@unistra.fr
\* Correspondence: agata.zubkiewicz@zut.edu.pl (A.Z.); spaszkiewicz@zut.edu.pl (S.P.); Tel.: +48-91-449-4589

**Abstract:** The influence of carbon multi-walled nanotubes (MWCNTs) and halloysite nanotubes (HNTs) on the physical, thermal, mechanical, and electrical properties of EVA (ethylene vinyl acetate) copolymer was investigated. EVA-based nanocomposites containing MWCNTs or HNTs, as well as hybrid nanocomposites containing both nanofillers were prepared by melt blending. Scanning electron microcopy (SEM) images revealed the presence of good dispersion of both kinds of nanotubes throughout the EVA matrix. The incorporation of nanotubes into the EVA copolymer matrix did not significantly affect the crystallization behavior of the polymer. The tensile strength of EVA-based nanocomposites increased along with the increasing CNTs (carbon nanotubes) content (increased up to approximately 40% at the loading of 8 wt.%). In turn, HNTs increased to a great extent the strain at break. Mechanical cyclic tensile tests demonstrated that nanocomposites with hybrid reinforcement exhibit interesting strengthening behavior. The synergistic effect of hybrid nanofillers on the modulus at 100% and 200% elongation was visible. Moreover, along with the increase of MWCNTs content in EVA/CNTs nanocomposites, an enhancement in electrical conductivity was observed.

## 1. Introduction

In recent years, polymer nanocomposites containing various nanofillers such as graphite nanoplatelets and carbon nanotubes as well as montmorillonite nanoclays have attracted enormous interest in academic and industrial fields. The unique properties of the polymer nanocomposites, such as flame retardancy, improved thermal stability, increased mechanical properties, and gas barrier properties, depend not only on the properties of nanofillers and polymer matrix but also on the interfacial contact and interactions between the nanofiller and the polymer matrix.

Among the vast range of nanofillers, one of the most promising are carbon nanotubes (CNTs). They were discovered in 1991 by Sumio Iijima [1]. CNTs can be broadly categorized as single-walled carbon nanotubes (SWCNTs), with a typical diameter between 1 and 2 nm, and multi-walled carbon nanotubes (MWCNTs), with an outer diameter between 3 and 30 nm or more, depending on the number of graphitic layers forming their structure [2]. They exhibit excellent mechanical (elastic modulus: 1 TPa–1.7 TPa), thermal (thermal conductivity higher than 3000 W·m$^{-1}$·K$^{-1}$), and electrical (electrical conductivity: $10^5$ S·m$^{-1}$–$10^7$·S·m$^{-1}$) properties [3]. Moreover, CNTs possess low mass density, and large aspect ratio (length to diameter ≈ 1000) [4]. Despite their numerous advantages, carbon nanotubes also suffer from a few drawbacks. Because of the nanometric dimensions, CNTs have a strong tendency to aggregate. Moreover, carbon nanotubes are relatively expensive. Despite this, CNTs have gained a great deal of research interest, for over 20 years, especially as a reinforcing nanofillers in polymer nanocomposites [5]. In particular, great attention has been focused on multiwalled carbon nanotubes (MWCNTs) as fillers in polymer materials, such as epoxy resins [6–8], polyethylene (PE) [9,10], polypropylene (PP) [11–13], polyurethanes (PU) [14,15], etc.

Another type of nanotubular structures is naturally occurring halloysite nanotubes (HNTs). They were reported for the first time in 1826 by Berthier [16]. HNTs are abundantly available nanoparticles formed by rolled kaolin sheets with chemical composition $Al_2Si_2O_5(OH)_4·2H_2O$ (the hydrated form with one layer of water in the interlayer spaces: HNTs-10 Å). In a dry climate, they can also occur in an anhydrous form (with an interlayer spacing of 7Å) with the formula $Al_2Si_2O_5(OH)_4$ [17]. HNTs are 1:1 phyllosilicates that have a hollow tubular morphology which results from the wrapping of silicate sheets, consisting of one tetrahedral and one octahedral sheet, that are connected through hydrogen bonding and weak Van der Waals interactions [18]. HNTs were found to occur in soils all over the world, but the most important deposits are located in the United States, New Zealand, and Poland [19]. The pure material is white, however, as a result of impurities from ferric ions, it may be slightly red [20]. Typically, the HNTs lengths range between 300 and 1500 nm, while their inner diameters are 15–100 nm, and outer diameter 40–120 nm [21,22]. HNTs are low-cost, and eco-friendly materials that can be more easily dispersed in a polymer matrix than carbon nanotubes [19]. Compared to other layered silicates, they are characterized by relatively low hydroxyl content on their outer surfaces. It is related to the fact that most of the aluminols are located inside the tubes. In the outer surface of the HNTs are located mainly siloxane groups [16,23]. Moreover, HNTs have quite a high aspect ratio (10–50), high resistance to heat and chemical substances, and due to the empty lumen structures, relatively low density (2.14–2.59 g/cm$^3$) [19]. The surface of HNTs is negatively charged (at pH > ~2), whereas the inner surface is charged positively [16]. Because of the fact that HNTs are naturally occurring and much cheaper, yet morphologically similar to multiwalled carbon nanotubes, the HNTs could be an alternative for more expensive CNTs for selected applications. A lot of research has been carried out on the nanocomposites based on HNTs with various polymer such as PE [24,25], PP [23,26,27], PA [22] epoxy resin [28,29]. They can provide a significant improvement in the thermal stability, fire resistance, and mechanical properties of composites.

A range of novel materials can be obtained by the simultaneous introduction of two types of nanofillers to the polymer matrix. Hybrid materials combine the properties of both fillers, and may also exhibit additional properties, because of the synergistic effects [30–34]. Recent research on the manufacturing of hybrid thermoplastic composites focuses on hybrid reinforcement in order to achieve better or tailored mechanical performance [35–39]. In the case of nano-reinforcements, this usually pertains to obtaining some distinctive physical properties e.g., electrical conductivity, thermal resistance, or barrier properties.

Ethylene vinyl acetate (EVA) is an important copolymer, widely used in various applications such as wire and cable insulations, shoe soles, noncorrosive layers, and component packaging. EVA has good flexibility, low cost, and good barrier properties. On the other hand, EVA has low tensile strength, thermal stability, and high flammability. All these disadvantages can be overcome by adding nanofillers to the polymer matrix. There are relatively fewer works describing the effect of HNTs' content on the mechanical and thermal properties of EVA. Suvendu Padhi et al. reported that HNTs could improve the mechanical properties and thermal stability of EVA [18]. Moreover, the addition of HNTs to EVA could enhance water resistance and oxygen permeability [40]. Various CNTs containing EVA-based nanocomposites had also been reported [41–45]. The mass production of high quality CNTs at lower cost and their exceptional electrical, thermal, and mechanical properties, make it one of the most attractive nanofillers. The addition of carbon nanotubes to the EVA matrix can improve mechanical, thermal, and electrical characteristics.

In this work, we manufactured and compared nanocomposites based on EVA copolymer containing MWCNTs, HNTs, and the mixture of both of them (hybrid system) in mass ratio 1:1. The morphology, thermal, mechanical, and electrical properties of the manufactured nanocomposites were characterized.

## 2. Materials and Methods

### 2.1. Materials

Ethylene vinyl acetate copolymer (EVA Elvax 40L-03, DuPont DuPont Company, Wilmington, DE, USA) containing 40 wt.% of vinyl acetate was applied as a matrix in the obtained nanocomposites. According to producer data, it has a density of 0.967 g/cm$^3$ and a melt flow rate of 3 g/10 min (at 190 °C and 2.16 kg). Halloysite nanotubes (HNTs) with a diameter of 30–70 nm, length of 1–3 μm, density of 2.53 g/cm$^3$, the pore size of 1.26–1.34 mL/g pore volume, and surface area of 64 m$^2$/g were obtained from Sigma-Aldrich. Multi-walled carbon nanotubes (MWCNTs, Nanocyl® NC7000™) were purchased from Nanocyl SA (Sambreville, Belgium). The nanotubes had an average diameter of 9.5 nm and a length of 1.5 μm, and the specific surface area of 250–300 m$^2$/g, the density of 1.75 g/cm$^3$, volume resistivity of 10$^{-4}$ Ω·cm, according to the supplier's specification.

### 2.2. Composite Manufacturing

#### 2.2.1. Compounding

The nanocomposites based on EVA containing MWCNTs, HNTs, or the hybrid system of MWCNTs/HNTs (1:1) were prepared by melt blending using a counter-rotating, tight intermeshing twin-screw extruder: Leistritz Laborextruder LSM30 (L/D = 23, D = 34 mm). The nanofillers and EVA granulate were fed separately using two gravimetric feeders into their feed section. The compounding was carried out at temperatures ranging from 50 to 115 °C from the feed section to the nozzle and at 40 rpm. The extruded strands of compounds were then cooled in a water tank and subsequently pelletized. Three series of nanocomposites with different content of nanofillers were prepared: nanocomposites containing MWCNTs with filler content of 2, 4, 6, and 8 wt.%, nanocomposites containing HNTs with filler content of 2, 4, 6, and 8 wt.%, and hybrid nanocomposites containing both MWCNTs and HNTs (at mass ratio 1:1) with the total fillers' content of 4 and 6 wt.%. EVA compounds with MWCNTs, so as with the mixture of MWCNTs with HNTs, were manufactured in one-step compounding. The compounds with HNTs in turn were manufactured in two-step compounding due to hindrance in the feeding of small portions of HNT nanoclay, which has high bulk density. For this purpose, the masterbatch of EVA/HNT 68/32 wt.% was manufactured and was subsequently diluted with EVA to the set filing ratios in the second compounding process. The set and obtained filling ratios of manufactured EVA compounds, as well as their corresponding volumetric filling ratios are presented in Table 1.

**Table 1.** Compositions and the injection pressure of ethylene vinyl acetate (EVA)-based nanocomposites.

| Material | Set Filler Content (wt.%) | Filler Real Content (wt.%) | MWCNTs Volumetric Content (vol.%) | HNTs Volumetric Content (vol.%) | Injection Pressure-Type A Samples (bars) | Injection Pressure-Conductivity Samples (bars) |
|---|---|---|---|---|---|---|
| EVA | - | - | - | - | 800 | 640 |
| EVA/2 wt.% CNT | 2 | 2.16 | 1.18 | - | 820 | 660 |
| EVA/4 wt.% CNT | 4 | 3.93 | 2.16 | - | 910 | 710 |

| EVA/6 wt.% CNT | 6 | 5.8 | 3.22 | - | 1040 | 910 |
| EVA/8 wt.% CNT | 8 | 7.94 | 4.45 | - | 1170 | 970 |
| EVA/2 wt.% HNT | 2 | 2.01 | - | 0.78 | 690 | 560 |
| EVA/4 wt.% HNT | 4 | 4.02 | - | 1.57 | 680 | 550 |
| EVA/6 wt.% HNT | 6 | 6.03 | - | 2.38 | 710 | 590 |
| EVA/8 wt.% HNT | 8 | 8.03 | - | 3.20 | 730 | 600 |
| EVA/4 wt.% CNT + HNT (1:1) | 4 | 3.97 | 1.09 | 0.78 | 810 | 660 |
| EVA/6 wt.% CNT + HNT (1:1) | 6 | 6.05 | 1.68 | 1.20 | 900 | 710 |

2.2.2. Injection Molding

Nanocomposites were dried before injection molding at 43 °C for ~12 h in a POL-EKO SLW115 oven (POL-EKO-APARATURA sp.j., Wodzislaw Slaski, Poland) with forced convection. The standard test specimens were injection molded using an ARBURG ALLROUNDER (ARBURG GmbH + Co KG, Lossburg, Germany) 270 S 350-100 (clamping force 350 kN, screw diameter 25mm, L/D = 20). Type A (dogbone) specimens for tensile testing were manufactured in accordance with EN ISO 294, while samples of dimensions 60 × 60 × 2 mm were manufactured for electrical conductivity measurements. The barrel temperatures were set to 100–150 °C from the first zone to the nozzle. Mold temperature was kept ~30 °C. In order to bring the processing of samples closer to conditions of extrusion, a relatively slow constant injection volume flow of 10 ccm/s was used during injection molding, which resulted in actual injection pressures presented in Table 1. Injection time was ~3.7 s, while the holding pressure was set to rise from 400 to 1000 bars for 15 s For EVA/HNT compounds the holding pressure was reduced to 800 bars. Backpressure by dosing was set to 30 bars. Cooling time was 30 s The whole injection molding cycle amounted to around minute. In turn, MWCNT's increased the melt viscosity, which was reflected by up to 50% higher injection molding pressures for the highest 8 wt.% filling ratio.

*2.3. Measurements*

The morphology of the nanocomposites was analyzed using a scanning electron microscope (FE-SEM, Hitachi SU-70, Naka, Japan). Before SEM analysis, the samples were cryofractured in liquid nitrogen and then coated with a thin film of palladium-gold alloy, using thermal evaporation PVD (physical vapor deposition) method to provide electric conductivity.

Transmission electron microscopy (TEM) was performed on a 2100 F Jeol microscope. Prior to the analysis, the samples were cooled down in liquid nitrogen and cut into thin-layered pieces.

The X-ray Diffraction (XRD) analysis of the EVA-based nanocomposites was conducted with the use of a Panalytical X'Pert diffractometer (Malvern Panalytical, Malvern, UK) operating at 40 V and 40 mA with CuK$\alpha$ radiation ($\lambda$ = 0.154 nm). The samples were scanned from $2\theta = 4°$ to $70°$ with a step of $0.02°$.

The Fourier Transform Infrared Spectroscopy (FTIR) spectra of the EVA-based nanocomposites were recorded on a Tensor-27 spectrophotometer (Brucker, Ettlingen, Germany), in the range of 4000–400 cm$^{-1}$. Measurements were carried out using the attenuated total reflectance (ATR) technique.

Differential scanning calorimetry (DSC) measurements of the obtained nanocomposites were conducted on a Mettler Toledo DSC1 instrument (Mettler Toledo GmbH, Greifensee, Switzerland).

The samples were heated up/cooled down under nitrogen flow with a heating rate of 10 °C/min in the temperature range −75 °C to 175 °C. The crystallinity degree ($X_c$) of the investigated materials was calculated using the following equation:

$$X_c = \Delta H_m / \Delta H_m^o (1 - \varphi_n) \tag{1}$$

where: $\Delta H_m^o$ (= 293 J/g) is the enthalpy change of melting for a fully crystalline PE [46], $\varphi_n$ is a weight content of nanofiller, and $\Delta H_m$ is derived from melting peak area on DSC thermogram.

Thermal and thermo-oxidative stability of prepared nanocomposites was determined by thermogravimetric analysis (TGA 92-16, Setaram). Samples were heated in nitrogen and oxidizing atmosphere ($N_2:O_2$ = 80:20 vol.%) from temperature range 20 °C to 700 °C.

Density ($d_R$) was measured at 22 °C, using hydrostatic scales (Radwag WPE 600C, Radom, Poland), calibrated using working standards of known density. For each sample, five measurements were conducted, and then the results were averaged to obtain a mean value. Theoretical densities of the obtained materials were calculated using the rule of mixture, taking into account the real density of EVA, filler densities according to the producer data, and their volume fractions.

The melt flow rate (MFR) was measured on a melt indexer (CEAST, Pianezza TO, Italy) at a temperature of 190 °C and under 2.19 kg load, according to ISO 1133 specification.

Hardness was measured using the Shore D apparatus (Zwick GmbH, Ulm, Germany) after 15 s of loading, according to a standard ISO 868. Ten measurements were conducted and then the results were averaged to obtain a mean value.

The tensile properties of the prepared nanocomposites were measured using Autograph AG-X plus universal testing machine (Shimadzu, Duisburg, Germany) equipped with a 1 kN Shimadzu load cell, a non-contact optical extensometer, and the TRAPEZIUM X computer software (version 1.00 provided by Shimadzu, Duisburg, Germany). Tests were performed at room temperature, with a strain rate of 250 mm/min to break. Tensile strength and elongation at break of the nanocomposites were determined. The reported values are the average values of ten measurements. Cyclic tensile measurements were performed using the same equipment, with a testing speed of 100 mm/min. The samples were stretched until the specified strain value was reached and then the tensile force was released to the zero. This procedure was repeated, with increasing deformation value, until the sample broke. The following strains were established for our test: 5%, 15%, 25% 50% 100%, 200%, and 400%.

The electrical conductivity of the obtained nanocomposites was evaluated by measurements of resistivity using Electrometer 6517A (Keithley Instruments, Inc., Germering, Germany) device together with a set of Keithley 8009. The measurements were performed accordingly to the standard PN-88/E-04405.

## 3. Results and Discussion

### 3.1. Dispersion State Investigation

The quality of the dispersion of filler in a matrix is a crucial factor that determines the final properties of composite material. The nanofillers dispersion was examined through scanning electron microscopy (SEM) and transmission electron microscopy (TEM). The representative micrographs of the nanocomposites containing 4 and 6 wt.% of nanofillers are shown in Figure 1. MWCNTs distribution appears to be rather homogenous in the entire polymer matrix, and no large agglomerates were observed (Figure 1a,b). However, from the comparison of EVA/CNTs and EVA/HNTs images, it is clear, that HNTs are more evenly distributed. CNTs tend to bundle together because of the van der Waals interaction between the individual nanotubes. In contrast to carbon nanotubes, HNTs have relatively low tube–tube interactions. It is related to the hydroxyl groups that are located on the HNTs surfaces. Moreover, the tubular morphologies with a relatively high aspect ratio limit the possibility of creating large-area contact between tubes [19]. Consequently, a very uniform dispersion of HNTs in the EVA matrix was obtained (Figure 1c,d). The SEM images of hybrid nanocomposites are presented in Figure 1e,f. Both nanofillers are well distributed within the polymer matrix. On the cryo-fractured surface of nanocomposites, the nanotubes pulled from the EVA copolymer matrix can be observed.

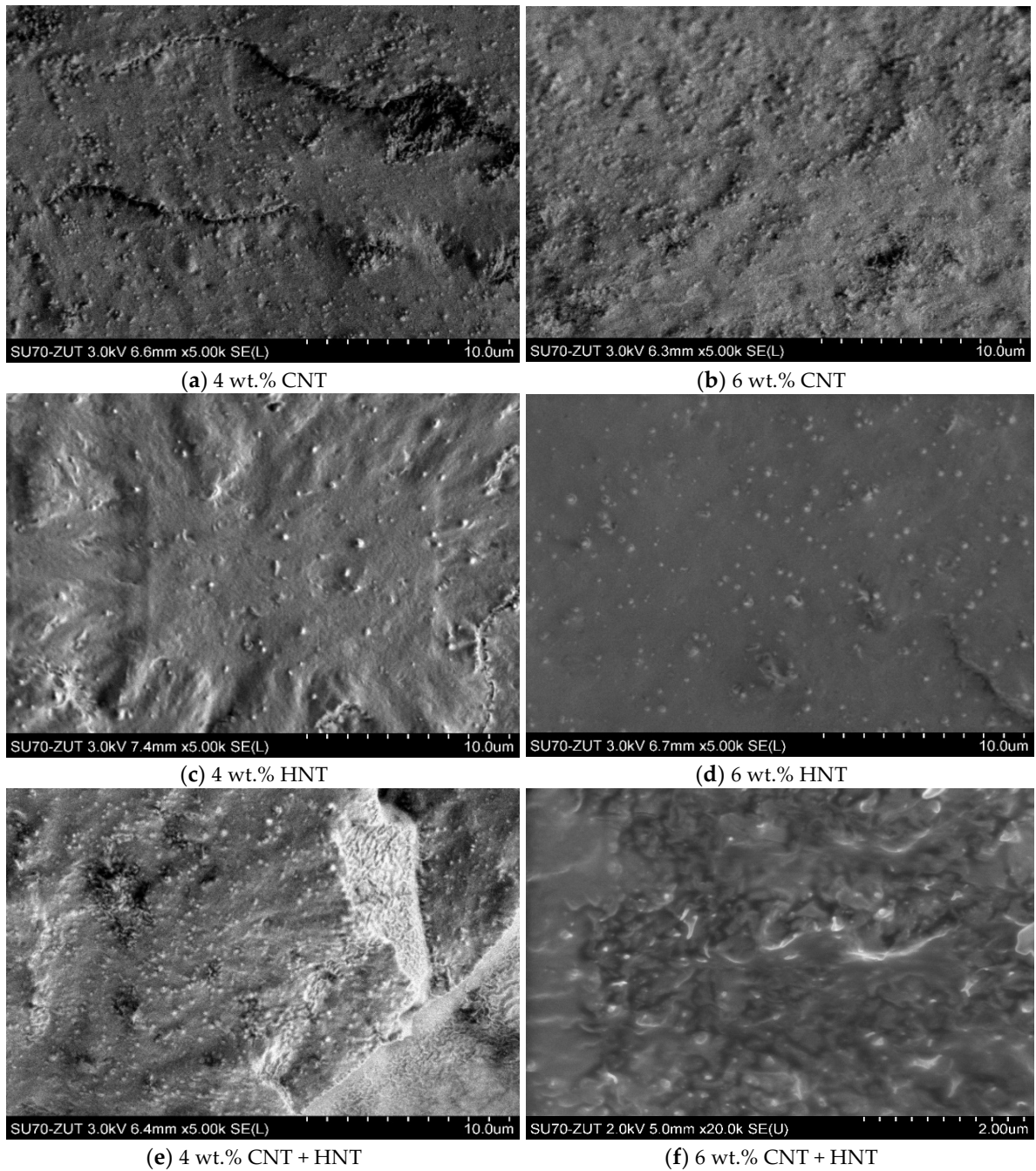

**Figure 1.** SEM images of EVA/multi-walled carbon nanotubes (MWCNT) (**a**, **b**), EVA/halloysite nanotubes (HNT) (**c**, **d**), and EVA/MWCNT + HNT (**e**, **f**) nanocomposites.

To confirm the state of dispersion of nanotubes in the matrix, TEM analysis for nanocomposite with the highest content of the MWCNTs and MWCNT/HNT hybrid system was also carried out. Figure 2 presents the TEM images of nanocomposites containing MWCNTs. Figures 3 and S1 (Supplementary Materials) represent the TEM images of EVA/6 wt.% CNT + HNT hybrid nanocomposite at different magnification. In Figure 3a, it can be seen that the dispersed halloysite nanotubes are relatively short. Probably, some of HNTs broke during compounding. Moreover, in contrast to the SEM observation, on TEM images a few agglomerates of nanotubes can be seen. At the higher magnification on TEM images (Figures 2 and S2), it can be seen that the individual and entangled agglomerates of CNTs were embedded in the EVA matrix.

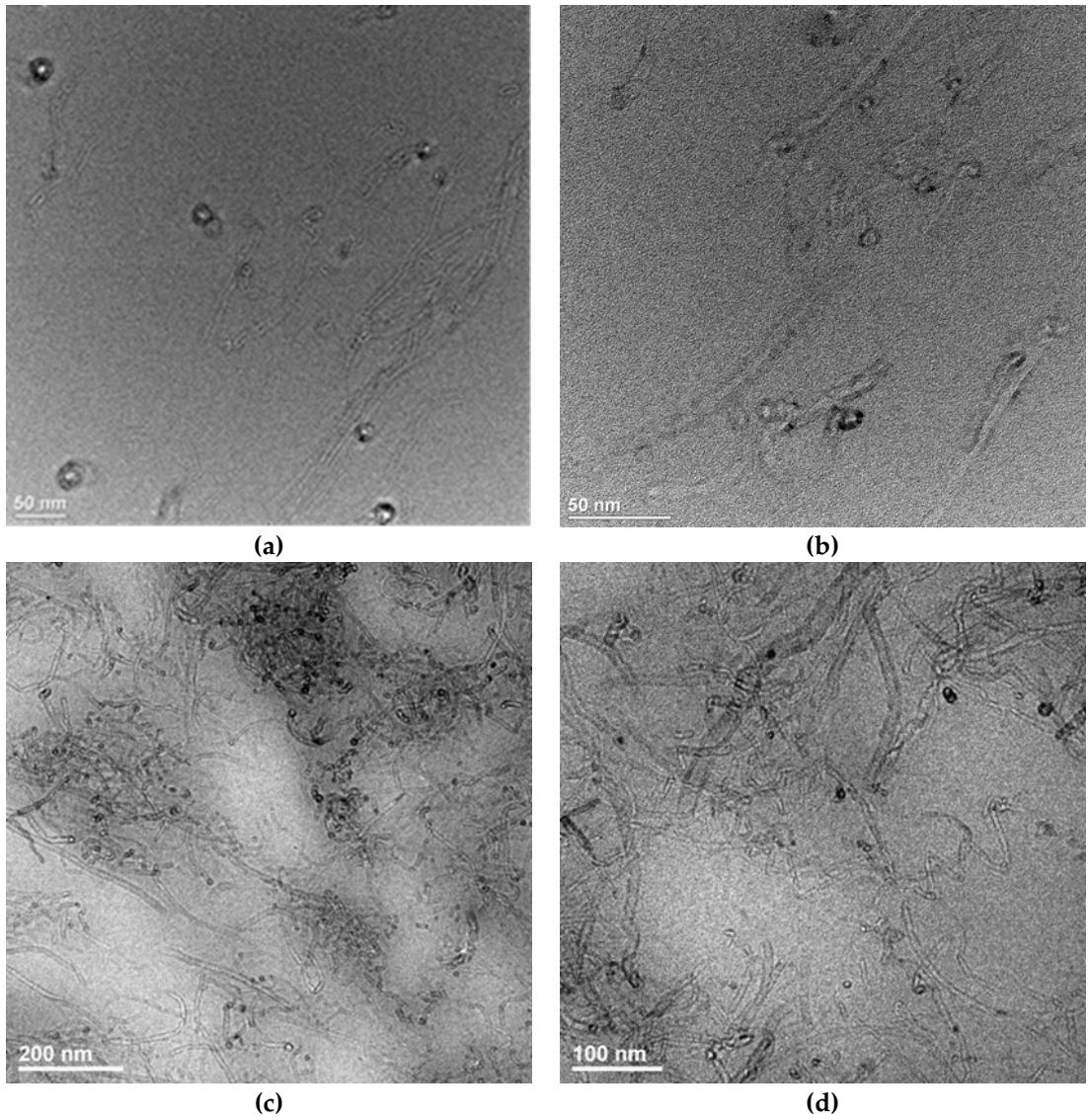

**Figure 2.** TEM images of EVA/6 wt.% CNT (**a**, **b**), EVA/8 wt.% CNT (**c**, **d**) nanocomposites at different magnification. Scale bars: 50 nm (**a**, **b**), 200 nm (**c**), 100 nm (**d**).

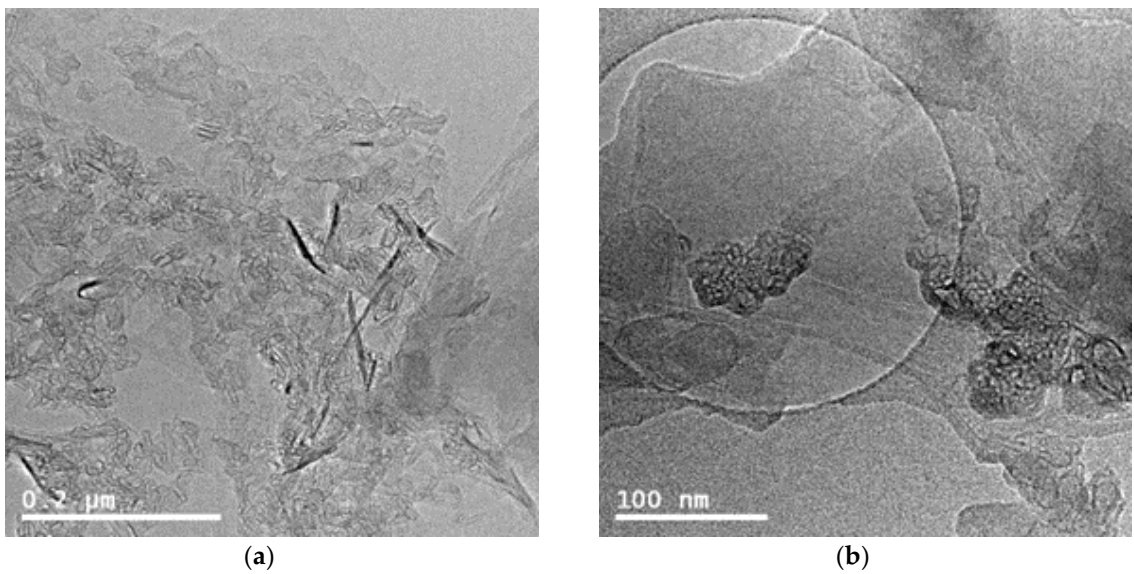

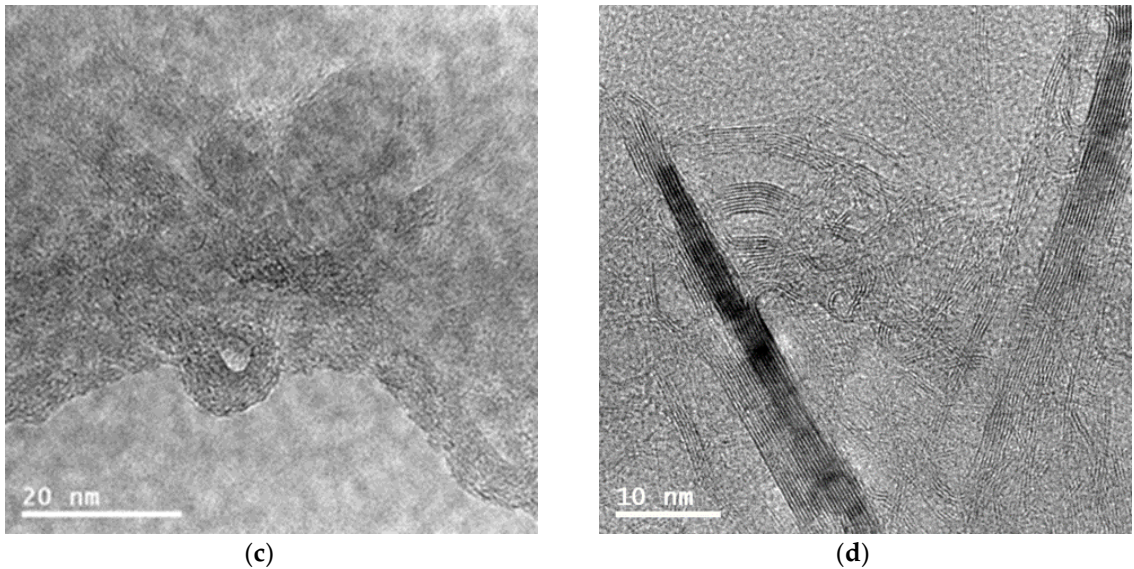

**Figure 3.** TEM images of EVA/6 wt.% CNT + HNT hybrid nanocomposite at different magnification. Scale bars: 0.2 μm (**a**), 100 nm (**b**), 20 nm (**c**), 10 nm (**d**).

*3.2. XRD Analysis*

The XRD patterns of the pure HNTs, MWCNTs powder, and EVA nanocomposites with their different loading are shown in Figure 4. The pristine HNTs shows a characteristic reflection peaks at diffraction angles (2θ) of 11.71, 19.99, 24.88, 26.63, 35.1 35.92, 38.31, and 62.40° corresponding to the d-spacings of 0.755 nm (001), 0.443 nm (020), 0.357 nm (110), 0.334 nm (113), 0.256 nm, (131), 0.234 nm (131)), 0.234 nm (133), and 0.148 nm (332), respectively [47]. The diffraction peaks at around 2θ = 12° and 2θ = 25° are attributed to the dehydrated form of HNTs, whereas the visible distinct peak at around 2θ = 63° indicates that the halloysite is a dioctahedral mineral [18,40]. For neat EVA, only a broad scattering reflection is found. It is located at around 2θ = 20.4° and indicates mainly an amorphous structure of the neat EVA matrix because of the low crystallinity of neat EVA (Table 2). It can be seen, that for EVA/HNTs nanocomposites, the HNTs characteristic peaks at 2θ of 12° and 25° are visible (Figure 4a) and thus, with the increase in the addition of HNTs, the intensity of these peaks increases. The position of diffraction peaks of EVA-based nanocomposites remained unchanged with different HNTs content. Figure 4b shows the XRD patterns of CNT, EVA/CNT, and hybrid nanocomposites. The characteristic peaks assigned to MWCNTs are seen at 2θ of 25.5° (d = 0.348 nm) and 42.7° (d = 0.211 nm), and they correspond respectively to the graphite indices of (002), which is related to d-spacing between graphene sheets, and (100), which is associated to the in-plane graphitic structure [48,49]. These characteristic peaks for MWCNTs were not present in XDR patterns for EVA/CNTs nanocomposites, because of the overlapping diffraction signals of CNTs and EVA and their low intensity (low MWCNTs concentration) in EVA matrix. The XRD patterns of EVA hybrid nanocomposites show a characteristic peak of HNTs at 2θ = 12.2°. The other diffraction peaks overlap with the peaks of EVA copolymer.

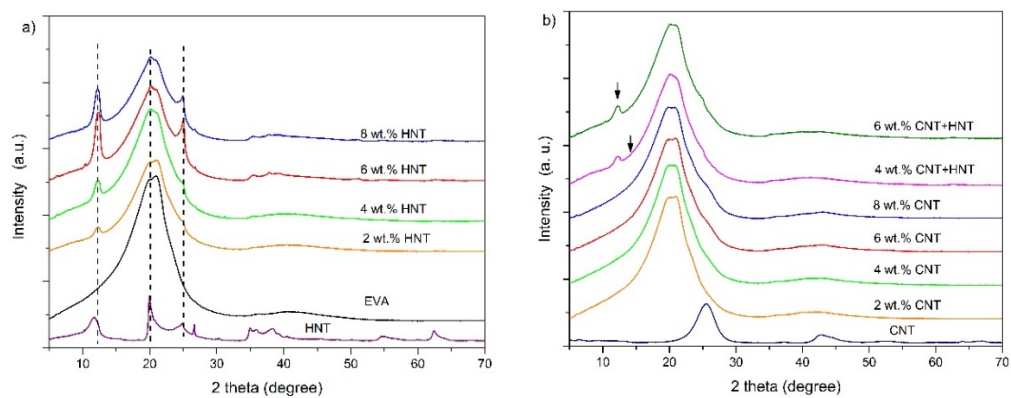

**Figure 4.** X-ray diffraction patterns of EVA/HNTs (**a**), EVA/MWCNTs, and MWCNT/HNT hybrid (**b**) nanocomposites.

**Table 2.** Phase transition temperatures, enthalpies of melting, and degree of crystallinity of EVA-based nanocomposites.

| Material | $T_g$ °C | $T_m$ °C | $\Delta H_m$ J/g | $T_c$ °C | $X_c$% |
|---|---|---|---|---|---|
| EVA | −20 | 50 | 5.79 | 26 | 1.98 |
| EVA/2 wt.% CNT | −18 | 51 | 5.88 | 24 | 2.00 |
| EVA/4 wt.% CNT | −17 | 51 | 5.29 | 23 | 1.81 |
| EVA/6 wt.% CNT | −16 | 50 | 5.39 | 23 | 1.84 |
| EVA/8 wt.% CNT | −14 | 50 | 5.06 | 22 | 1.73 |
| EVA/2 wt.% HNT | −18 | 51 | 5.78 | 26 | 1.97 |
| EVA/4 wt.% HNT | −18 | 51 | 5.35 | 26 | 1.83 |
| EVA/6 wt.% HNT | −18 | 51 | 5.32 | 26 | 1.82 |
| EVA/8 wt.% HNT | −17 | 51 | 4.60 | 26 | 1.57 |
| EVA/4 wt.% CNT + HNT | −17 | 52 | 5.37 | 23 | 1.83 |
| EVA/6 wt.% CNT + HNT | −17 | 51 | 5.61 | 23 | 1.92 |

$T_g$—glass transition temperature; $T_m$—melting temperature; $\Delta H_m$—enthalpy of melting; $T_c$—crystallization temperature; $X_c$—degree of crystallinity.

*3.3. Thermal Properties of the EVA Nanocomposites*

The DSC thermograms for EVA/CNT nanocomposites recorded during first (dashed line) and second heating, as well as cooling, as plotted in Figure 5a,b, respectively. Similarly, the DSC thermograms for the series of materials containing HNTs are presented in Figure 5c,d. Besides, in Figure 5e,f, the DSC thermograms were plotted for the samples containing 4 wt.% (Figure 5e) and 6 wt.% (Figure 5f) (in total) of CNT, HNT, and the mixture of both (CNT + HNT) at a mass ratio of 1:1. Likewise, the DSC parameters are summarized in Table 2. For the series of EVA-based nanocomposites containing CNTs, one observed that along with an increase of the CNTs' concentration, the value of the glass transition temperature ($T_g$) of EVA also increases. The melting temperatures ($T_m$) of nanocomposites in comparison to the neat EVA copolymer were comparable to one another, while a slight decrease in the crystallization temperature ($T_c$) was observed. EVA copolymer containing 40% by mass of VA has a low crystallinity of around 2%. The addition of MWCNTs and HNTs did not significantly affect the degree of crystallinity of nanocomposites, their degree of crystallinity is around 1.6–2.0%. Generally, CNTs have been proved to be good nucleating agents for polymer crystallization [50,51]. It has been reported that at low loading of CNTs in some semicrystalline polymers in the molten state they can induce crystallization at higher temperatures through decreasing the nucleation activation energy and increasing the nucleation density, leading to the acceleration of crystallization and the decrease of spherulites diameters simultaneously [52,53]. It was also found that the CNTs in some polymer systems can generate anti-nucleation effects, and super-nucleation effects on polymer matrices [54,55]. No effect of CNTs on the nucleation of polymer crystals has also been reported in some cases [56,57]. More recently, LDPE/CNTs nanocomposites have been prepared in our research group [58]. Our results show that crystallization behavior of PE in LDPE/CNT nanocomposites was not influenced by the presence of CNTs. However, herein in EVA nanocomposites, rather antinucleating behavior of CNTs was observed. As can be seen in Table 2, the crystallization temperature ($T_c$) of EVA nanocomposites decreases with the increase of CNTs content. Similar as in LDPE/CNTs nanocomposites, it can be a result of a low value of surface energy and a poor wettability of CNTs. This means that it can be difficult for them to induce aggregation of polymer chains on their surfaces [58]. In turn, the increase in $T_g$ results from the fact that CNTs may prevent the mobility of the copolymer chains, leading to an increase in $T_g$ [50]. For the series of EVA-based nanocomposites containing HNTs, similarly as in the case of the CNTs, the values of $T_m$ and $T_c$ for nanocomposites are comparable to the value observed for neat EVA copolymer. However, along with an increase in the concentration of HNTs, the value of $T_g$ increased. In some polymer systems, halloysite nanotubes act as nucleating agents accelerating crystallization rate [26]. The introduction of HNTs into EVA copolymer did not influence $T_c$ and the degree of nanocomposites crystallinity in comparison to the neat EVA copolymer. Therefore, one can deduce that HNTs rather frustrates chain ordering and mobility, than the crystallinity behavior itself [40]. In turn, for two hybrid nanocomposites containing 4 wt.% and 6 wt.% of CNTs and HNTs, the observations on the phase transition temperatures are comparable to the ones observed for EVA/CNTs and EVA/HNTs nanocomposites, i.e., comparable values of $T_m$, slightly lower values of $T_c$ and $x_c$, and increase in $T_g$. These values appear to be the result of the impact of both types of nanoparticles, without the apparent effect of any of the used ones.

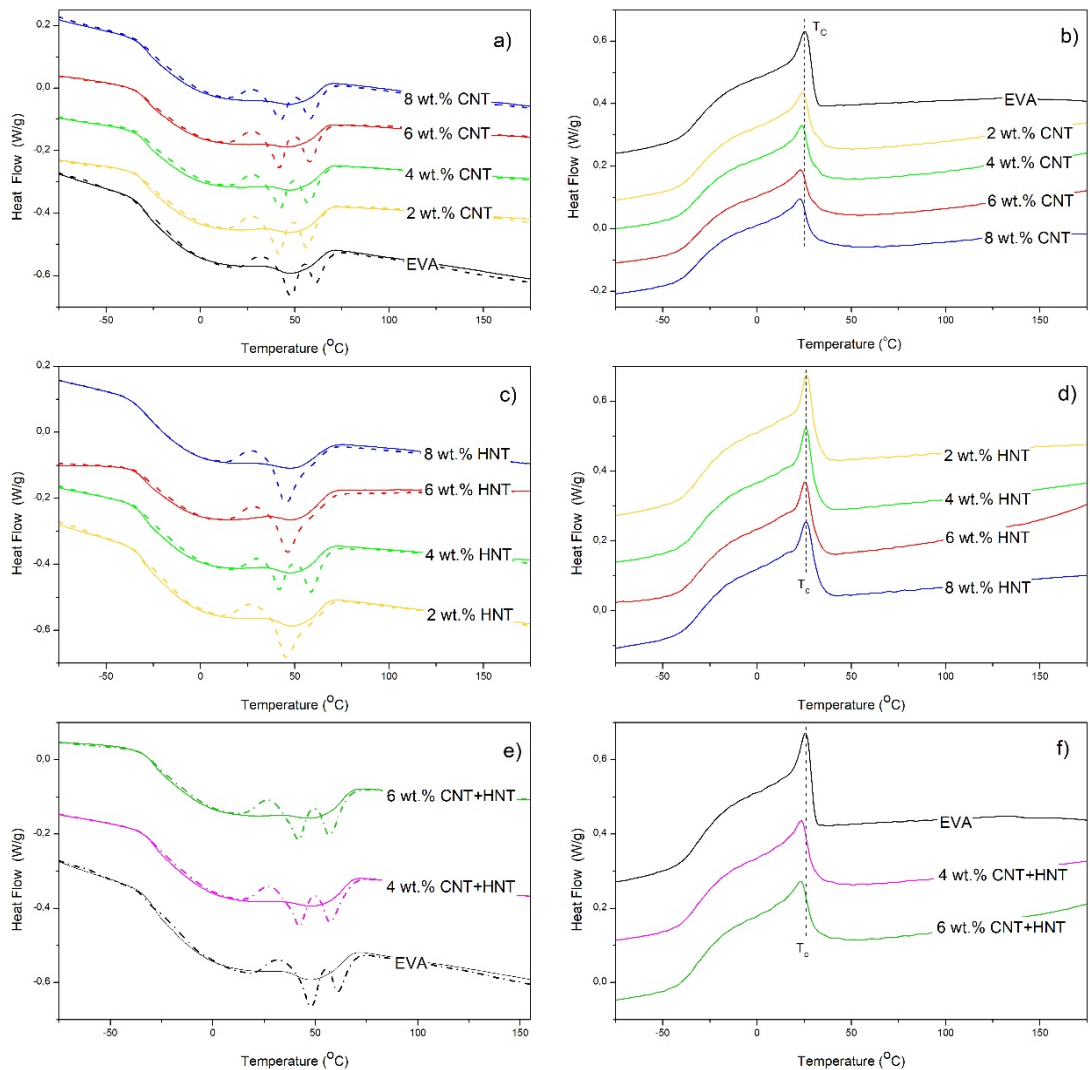

**Figure 5.** DSC thermograms for EVA/MWCNT nanocomposites (**a**, **b**), EVA/HNT nanocomposites (**c**, **d**), and hybrid nanocomposites (**e**, **f**).

Additionally, since the addition of carbon nanofillers, like CNTs, mineral nanotubes like HNTs or mineral nanoclays like MMT, can improve thermal and thermo-oxidative stability of polymer matrices, the mass loss and derivative mass loss curves for the series of materials have been presented in Figures 6 and 7. Moreover, in Table 3, the temperatures corresponding to 5, 10, and 50% mass loss and the temperatures at a maximum of mass-loss rate for the EVA-based nanocomposites have been tabulated. It is well-known that the thermal stability of polymer composites plays a crucial role in determining their processing and applications since it affects the final properties of the materials such as the upper limit usage temperature and dimensional stability [59]. It is now well accepted that the improvement in thermal stability of polymer nanocomposites containing CNTs is due to the following factors: barrier effect, the thermal conductivity of CNTs, physical or chemical adsorption, radical scavenging action, and polymer-nanotube interaction. For each polymer-CNT composite, thermal stability may be due to one mechanism or the combined action of several processes, which depend on the different components, microstructures, and exterior conditions [59]. Similarly, also HNTs are found to be a promising additive for improving the thermal stability and flame retardancy of polyolefins and other polymers [19]. The improvement of thermal stability and flame retardancy resulted from the barrier properties of HNTs combined with an encapsulation process of the polymer's degradation products inside the HNT lumens [60]. What is also important, because of their structure and chemical character, HNTs can be more easily dispersed into polyolefin matrices in comparison with other nanofillers, such as montmorillonite (MMT) [16,19,61]. This is a crucial factor for obtaining nanocomposites with better mechanical properties, higher thermal stability, and reduced flammability. Of course, there are studies on nanocomposites based on EVA with the addition of MWCNT [62] to improve the thermal stability, however, so far nobody has studied how a hybrid system of carbon nanotubes and mineral nanotubes can improve both thermal and thermo-oxidative stability. Herein we can see that in general, in the oxidizing atmosphere (Figure 6), a three-step degradation process is observed for all samples, while in an inert atmosphere a two-step degradation process is observed (Figure 7). The TGA thermograms of the EVA show that the first degradation process starts at 250 °C ($T_{onset}$) and is completed at 390 °C [63]. This process corresponds to the loss of acetic acid [64]. The second step corresponds to the degradation of the polyethylene chains and starts at approximately 410 °C and ends at 465 °C [65]. While the third stage starts (observed only in the oxidizing atmosphere) at approximately 480 °C and ends at 570 °C, in the case of neat EVA, whereas in the case of all composites, it starts at approximately 580 °C and ends at 635 °C. In the case of EVA/HNTs nanocomposites and EVA/CNTs+HNTs hybrid nanocomposites, this third step corresponds to the decomposition of carbonaceous-silicate char [66,67]. While in the case of EVA/CNT nanocomposites this third step corresponds to the decomposition of residue formed in the second step of decomposition. Additionally, in an oxidizing atmosphere, the first two characteristic temperatures ($T_{5\%}$ and $T_{10\%}$) were slightly reduced by the incorporation of HNTs, CNTs, and the mixture of both, while at the $T_{50\%}$ (i.e., at higher temperatures) one can observe the enhancement of thermo-oxidative stability by the incorporation of nanofillers, of even 20 °C for EVA/4 wt.% CNT + HNT. In the case of the analysis conducted in the inert atmosphere, almost all samples exhibited improvement in thermal stability. Only in the case of nanocomposites containing 2 and 4 wt.% of HNTs no improvement was observed. In general, the studies conducted on polymer nanocomposites containing HNTs [27,68,69] explained such deterioration of thermal stability by the physical adsorption of water on the external surface of HNTs, contributing to the degradation of polymer matrix [68]. In turn, Bidsorkhi et al. [40] demonstrated that the improvement in thermal stability is attributed to the homogeneous dispersion of HNTs, which originated from strong hydrogen bonding between surface functional groups of HNTs and vinyl acetate groups of EVA. In the case of obtained EVA/HNT nanocomposites, the interaction between HNTs and EVA were confirmed by FTIR spectroscopy. Figure 8 shows FTIR spectra of the used neat EVA copolymer for nanocomposite preparation. The characteristic absorption peaks at 1238 cm$^{-1}$ and 1735 cm$^{-1}$ are assigned to the C–O and C=O stretching in vinyl acetate, whereas reflections at 2850 cm$^{-1}$ and 2918

cm$^{-1}$ are attributed to the C-H stretching vibration in ethylene. For EVA nanocomposites with the content of 6 and 8 wt.% of HNTs, the peak corresponding to stretching vibrations of C=O of EVA shifts to lower wavenumber, from 1735 to 1726 and 1723 cm$^{-1}$, respectively. This shift may be attributed to hydrogen bonding interactions between the carbonyl groups (C=O) of EVA and the hydroxyl groups of HNTs.

**Table 3.** Temperatures corresponding to 5, 10, and 50% mass loss and the temperatures at a maximum of mass loss rate for the EVA-based nanocomposites obtained in an oxidizing and inert atmosphere.

| Sample | Air | | | | | |
|---|---|---|---|---|---|---|
| | $T_{5\%}$, °C | $T_{10\%}$, °C | $T_{50\%}$, °C | $T_{DTG1}$, °C | $T_{DTG2}$, °C | $T_{DTG3}$, °C |
| EVA | 320 | 331 | 441 | 351 | 450 | 535 |
| EVA/2 wt.% CNT | 314 | 328 | 453 | 346 | 488 | 609 |
| EVA/4 wt.% CNT | 314 | 330 | 457 | 348 | 486 | 609 |
| EVA/6 wt.% CNT | 309 | 328 | 459 | 346 | 485 | 609 |
| EVA/8 wt.% CNT | 311 | 330 | 454 | 348 | 483 | 602 |
| EVA/2wt.% HNT | 313 | 327 | 449 | 350 | 464 | 526 |
| EVA/4wt.% HNT | 310 | 325 | 447 | 346 | 461 | 526 |
| EVA/6 wt.% HNT | 308 | 327 | 450 | 344 | 464 | 525 |
| EVA/8wt.% HNT | 311 | 330 | 451 | 350 | 463 | 526 |
| EVA/4 wt.% CNT + HNT | 315 | 332 | 462 | 351 | 490 | 526 |
| EVA/6 wt.% CNT + HNT | 310 | 329 | 459 | 351 | 488 | 615 |
| Sample | Argon | | | | | |
| | $T_{5\%}$, °C | $T_{10\%}$, °C | $T_{50\%}$, °C | $T_{DTG1}$, °C | $T_{DTG2}$, °C | - |
| EVA | 322 | 336 | 448 | 349 | 470 | - |
| EVA/2 wt.% CNT | 330 | 341 | 456 | 350 | 473 | - |
| EVA/4 wt.% CNT | 327 | 340 | 462 | 352 | 477 | - |
| EVA/6wt.% CNT | 327 | 340 | 458 | 349 | 474 | - |
| EVA/8 wt.% CNT | 327 | 341 | 463 | 350 | 476 | - |
| EVA/2 wt.% HNT | 320 | 334 | 452 | 351 | 471 | - |
| EVA/4 wt.% HNT | 320 | 335 | 450 | 350 | 468 | - |
| EVA/6 wt.% HNT | 326 | 340 | 456 | 350 | 474 | - |
| EVA/8wt.% HNT | 325 | 339 | 456 | 351 | 473 | - |
| EVA/4 wt.% CNT + HNT | 324 | 338 | 455 | 351 | 473 | - |
| EVA/6 wt.% CNT + HNT | 333 | 344 | 461 | 351 | 475 | - |

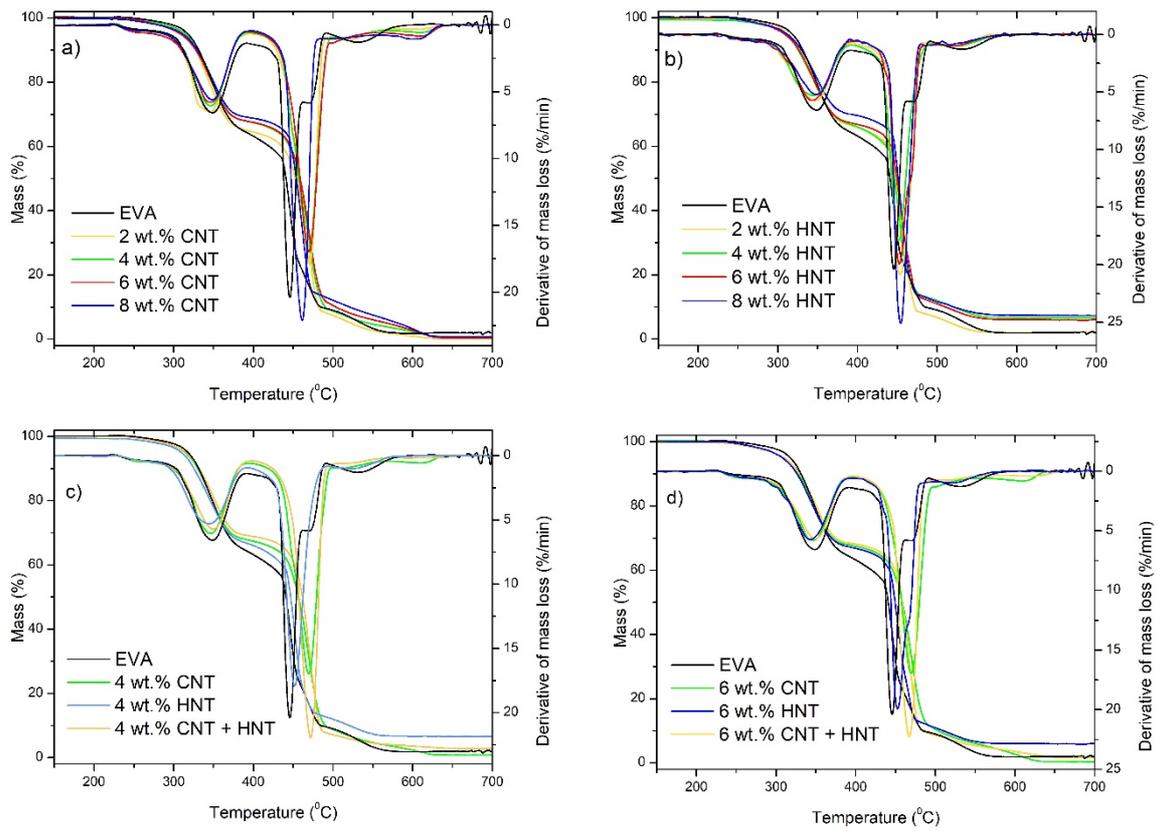

**Figure 6.** TG and DTG curves of: EVA/MWCNT nanocomposites **a**); EVA/HNT nanocomposites **b**); EVA-based nanocomposites at the total nanofillers content of 4 wt.% **c**); and EVA-based nanocomposites at the total nanofillers content of 6 wt.% **d**) measured in an oxidizing atmosphere.

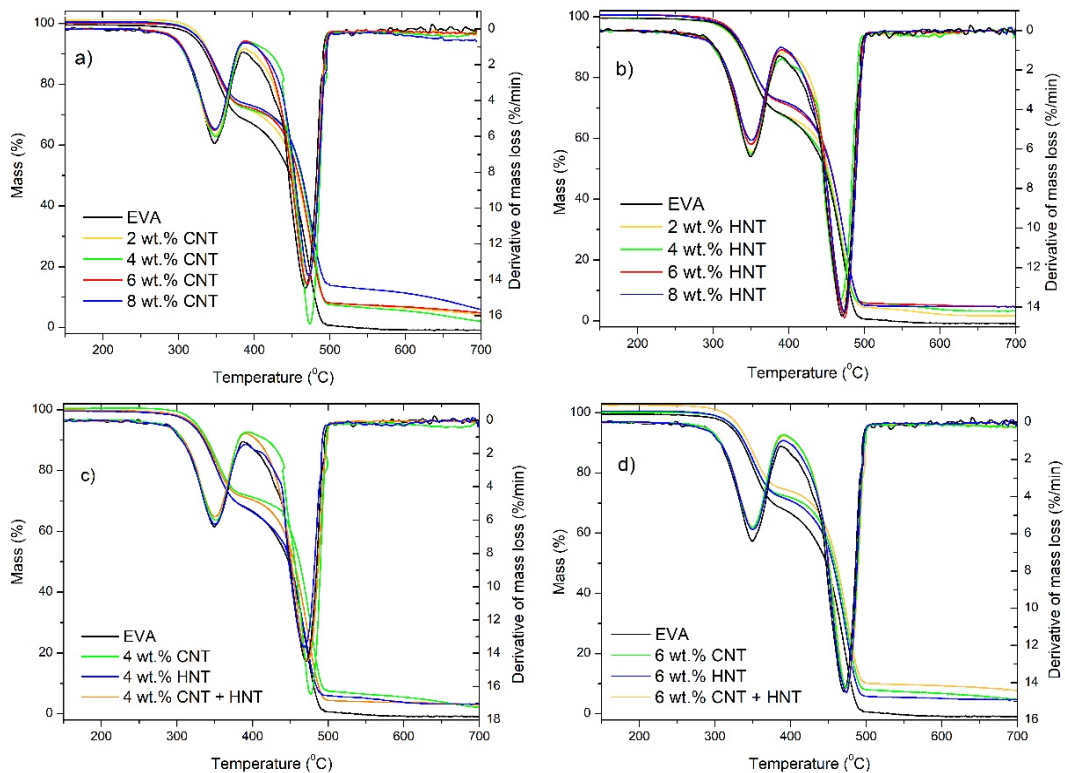

**Figure 7.** TG and DTG curves of: EVA/MWCNTs nanocomposites **a**); EVA/HNTs nanocomposites **b**); EVA-based nanocomposites at the total nanofillers content of 4 wt.% **c**); and EVA-based nanocomposites at the total nanofillers content of 6 wt.% **d**) measured in an inert atmosphere.

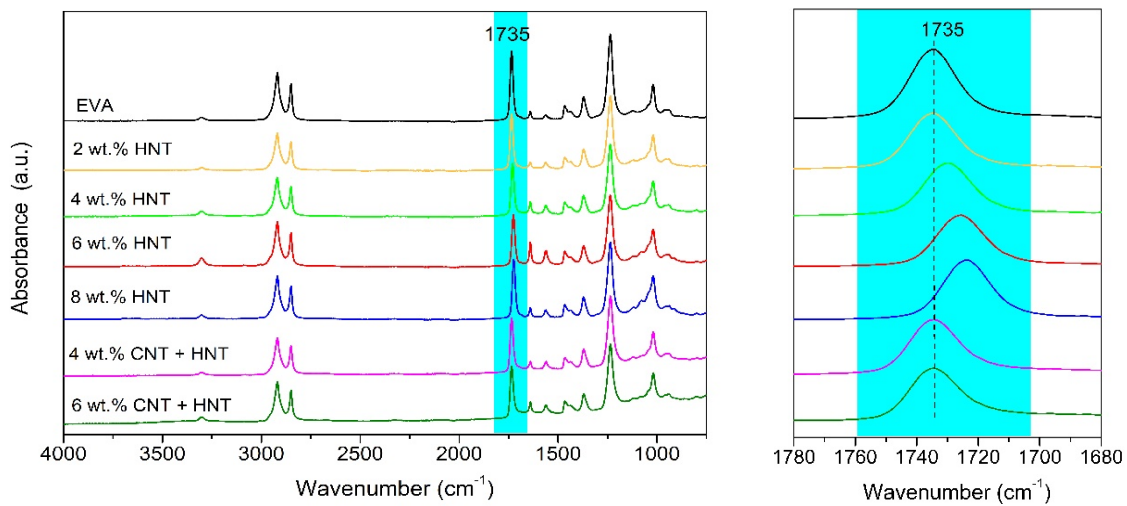

**Figure 8.** FTIR spectra for the neat EVA copolymer and EVA-based nanocomposites with HNTs and hybrid CNT + HNT.

As a result of the uniform dispersion of HNTs, the highest possible value of the surface-to-volume ratio for the nanofillers is achieved. Therefore, the degraded and/or degrading products of EVA polymer were entrapped within the tubular rods of HNTs, consequently contributing to a delay in the polymer decomposition process. Another possible interpretation for the improvement in the thermal stability of the nanocomposites could be the insulation effect of HNTs. Generally, layered silicates are thought to be an excellent thermal barrier that can effectively protect the matrix from being exposed to heat flow and thermal energy [70]. Although the improvement in thermo-oxidative and thermal stability of polymer/carbon nanotubes and polymer/layered inorganics-based nanocomposites has been reported extensively, the mechanism of such effect is still not yet well understood. Especially when two types of nanofillers, that differ in the properties and structure, are mixed. Herein, both hybrids exhibited a very decent improvement in thermal stability, even though no synergistic effect of property improvement was observed in this case. Generally, the most common explanation suggests that the enhancement in thermal stability derived from the mass and heat transfer barrier caused by a carbonaceous char (CNT-based composites) and carbonaceous-silicate char (HNTs-based composites) on the surface of the polymer melt [59,66,67]. Moreover, in the case of HNTs-nanocomposites, recent studies also suggest that the effect may be associated with a chemical interaction between the polymer matrix and the outer layer surface during thermal degradation and combustion processes [71].

*3.4. Physical Properties of EVA-Based Nanocomposites*

Physical properties of EVA-based nanocomposites such as density, hardness, melt flow rate, and some mechanical properties are submitted in Table 4 and Figures 9–11. The theoretical and real densities of the nanocomposites were found to increase with nanofillers content because of the higher density of nanotubes over the neat EVA. One can observe that slight deviations of the measured values from the theoretical densities, especially when HNTs were used as filler, were visible. However, the differences are relatively small, and the ratio of experimental density to theoretical density does not exceed 98%. Therefore, it can be concluded that the desired compositions were obtained.

**Table 4.** Density, hardness, and melt flow rate of EVA/MWCNTs + HNTs nanocomposites.

| Material | $d_t$ (g/cm$^3$) | $d_R$ (g/cm$^3$) | H (ShD) | MFR (g/10 min) |
|---|---|---|---|---|
| EVA | 0.967 | 0.973 ± 0.001 | 22.4 ± 1.7 | 2.99 ± 0.46 |
| EVA/2 wt.% CNT | 0.983 | 0.980 ± 0.001 | 26.3 ± 1.3 | 0.99 ± 0.13 |
| EVA/4 wt.% CNT | 0.991 | 0.989 ± 0.001 | 27.9 ± 1.1 | 0.22 ± 0.08 |
| EVA/6 wt.% CNT | 1.000 | 0.999 ± 0.001 | 30.4 ± 1.0 | 0.12 ± 0.05 |
| EVA/8 wt.% CNT | 1.010 | 1.009 ± 0.002 | 33.2 ± 1.4 | - |
| EVA/2 wt.% HNT | 0.986 | 0.982 ± 0.001 | 22.8 ± 1.0 | 2.74 ± 0.32 |
| EVA/4 wt.% HNT | 0.998 | 0.989 ± 0.004 | 23 ± 1.2 | 4.05 ± 0.27 |
| EVA/6 wt.% HNT | 1.011 | 0.997 ± 0.001 | 23.1 ± 1.0 | 3.76 ± 0.46 |
| EVA/8 wt.% HNT | 1.024 | 1.008 ± 0.001 | 24.5 ± 1.0 | 3.90 ± 0.18 |
| EVA/4 wt.% CNT + HNT | 0.995 | 0.991 ± 0.001 | 25.6 ± 0.7 | 1.87 ± 0.14 |
| EVA/6 wt.% CNT + HNT | 1.006 | 1.001 ± 0.001 | 25.6 ± 1.5 | 1.22 ± 0.23 |

$d_t$—theoretical density; $d_R$—density real; H—Shore hardness, scale D; MFR—melt flow rate at 190 °C and 2.16 kg.

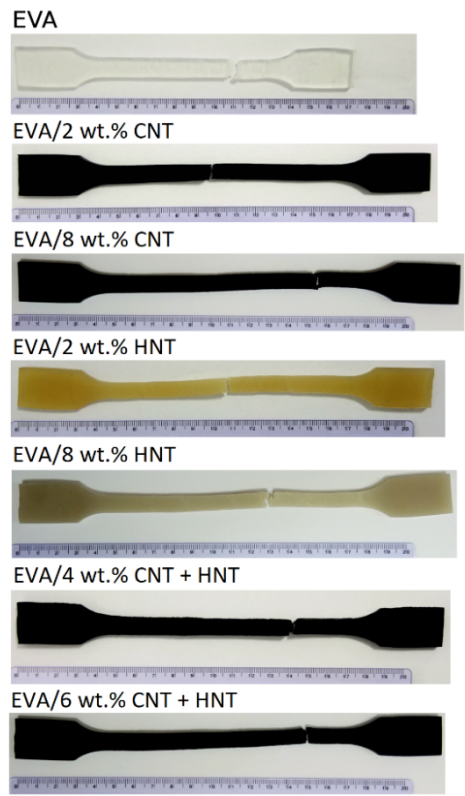

**Figure 9.** Samples of EVA-based nanocomposites after stretching.

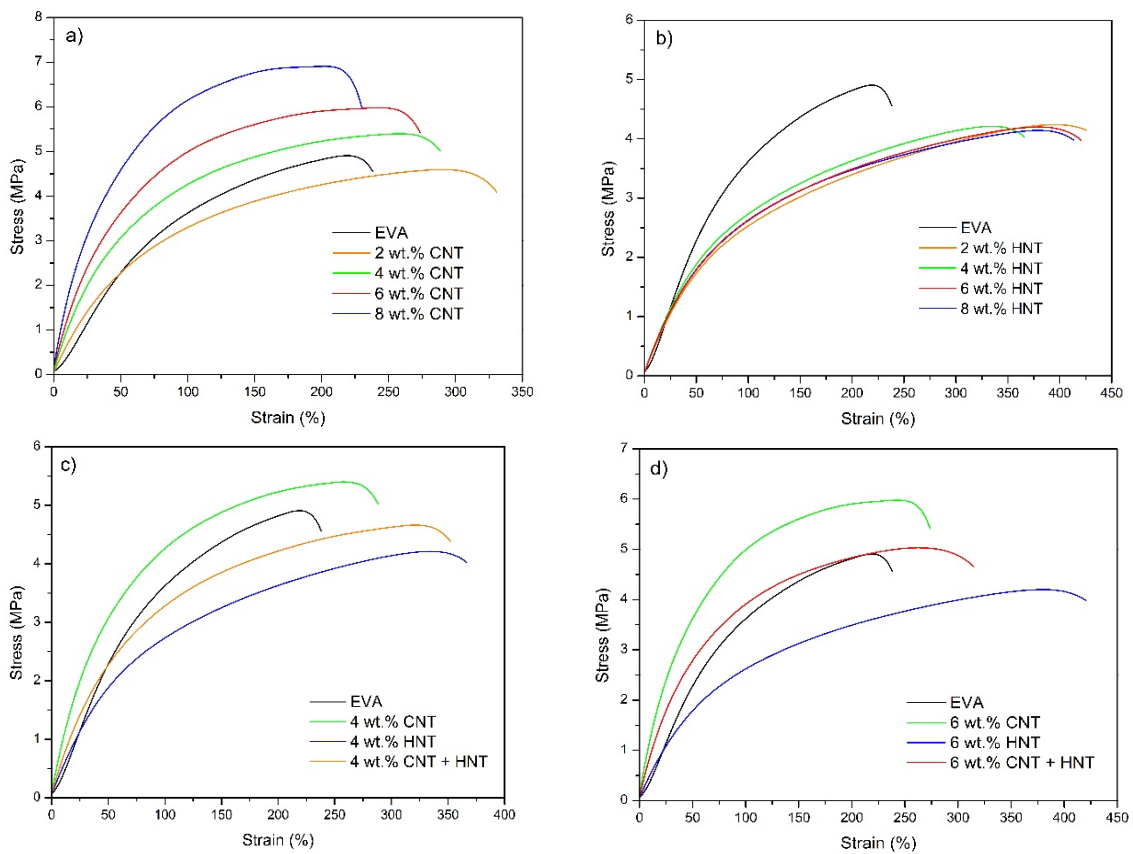

**Figure 10.** Stress–strain curves of EVA/MWCNT nanocomposites **a**), EVA/HNT nanocomposites **b**), EVA-based nanocomposites at the total nanofillers content of 4 wt.% **c**), and EVA-based nanocomposites at the total nanofillers content of 6 wt.% **d**).

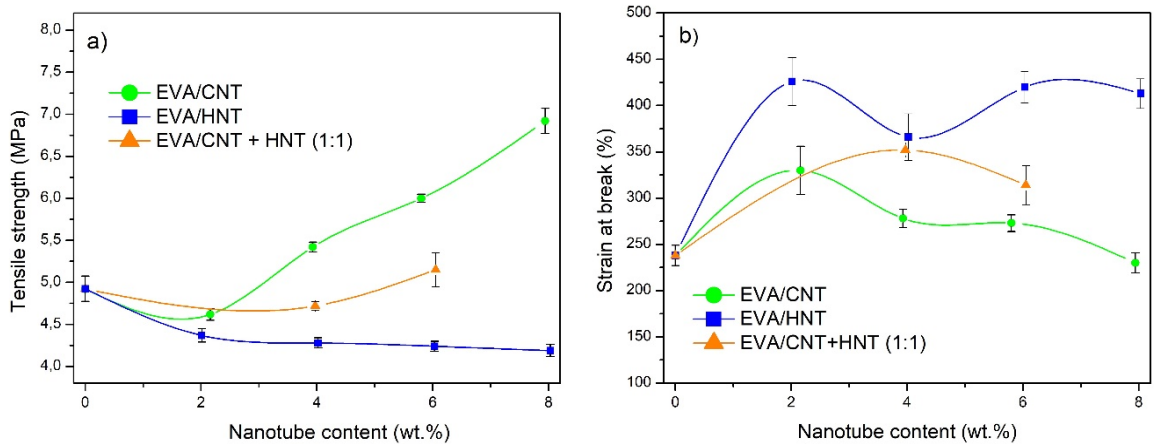

**Figure 11.** Tensile strength **a**) and strain at break **b**) of EVA/MWCNTs + HNTs nanocomposites with respect to nanofillers content.

The hardness test results show an increase in the Shore D hardness values with increasing fillers content (Table 4). Especially the presence of rigid reinforcement MWCNTs cause a considerable increase in nanocomposites hardness, which for EVA/CNT nanocomposite at a MWCNTs loading of 8 wt.% is approximately 48% higher than those of the neat EVA copolymer, while the hardness of the nanocomposite with 8 wt.% of HNTs is about 9% higher. In turn, as expected, the hybrid nanocomposites have intermediate hardness values fitting between nanocomposites filled with MWCNTs and HNTs.

Since MWCNTs are characterized by high Young's modulus (~1 TPa) and aspect ratio their introduction in the EVA matrix increases the composite stiffness, which can be observed in Figure 10 a, as a steepening of stress–strain curves. On the other hand, the filling of EVA with HNT's reduces the stiffness (Figure 10b). The tensile strength of EVA composites increases with the higher filling ratio of MWCNTs, but in case of filling with HNTs the strength is diminished and the increase of filling ratio has little influence on the tensile performance of the composite (Figure 11a). The addition of MWCNTs reinforces the EVA matrix, while HNT filler gives an adverse effect in terms of composite stiffness and tensile strength. However, the HNTs increase to a great extent the strain to break of EVA. The increase of strain at break and a slight decrease of tensile strength is also noticeable at 2 wt.% of MWCNTs filling ratio, but the effect is explained by high stiffness of MWCNTs, therefore it diminishes with higher filling ratios of MWCNTs. Taking into account that HNTs are considered to be fibrous nanoparticles, however with lower aspect ratio and lower stiffness than their carbon counterpart, they are still much stiffer than the neat EVA copolymer. Therefore, the increase of strain to break and reduction of strength and stiffness by filling with HNTs may be an effect of HNTs affecting the supermolecular structure of EVA. This is also relevant to MWCNTs although this effect is covered up by greater gains in terms of reinforcement at filling ratio > 2 wt.%.

The polymer crystal phase transformation or polymer chain intercalation leads usually to such effect and has been recorded for some polymer/nanoparticle systems if the good distribution of nanoparticles in the polymer chains is provided, which can be obtained only by good compatibility between the matrix and particles [72,73]. The increase of strain at break was usually obtained by incorporation of layered silicates in such polymers as polyvinylidene fluoride (PVDF), ethylene propylene diene rubber (EPDM), and polyurethanes (PU) [72,74-76]. Among them also PU nanocomposites with HNTs gave an extraordinary coupled increase of strength and strain to break [33]. Also, layered graphene introduced in-situ in polyester altered the same way the stress-strain behavior [77]. This plasticizing effect is also apparent by 15% lower injection molding pressures for the EVA/HNT compounds in comparison to native EVA (Table 1).

The extent degradation of the EVA matrix upon processing may be ruled out as this would also decrease strains to break of manufactured composites. The addition of HNT's increases also the flow of EVA, which was reflected indirectly in the results MFR presented in Table 4 (MFR). However, it cannot be assessed whether this was caused by EVA/HNTs interaction, degradation of EVA, or a combination of both factors. The interaction of HNTs and MWCNTs particles with EVA polymer chains has not been observed in the results of DSC measurements (Table 2), where only minute changes in the enthalpies of crystal melting and their characteristic temperatures were recorded. Nevertheless, in the EVA nanocomposites, where significant alternation of the supermolecular structure was proven, also, changes have not been observed in the DSC measurements [26,65]. It is also important to notice that MWCNTs and HNTs limited the recoiling of the polymer chains as the samples were not returning to their initial dimensions as quickly as for the neat EVA (Figure 9).

To investigate the elastic deformability and reversibility of the obtained nanocomposites, cyclic tensile tests were carried out. Results are presented in Figures 12–14. The contours made by the loops are consistent with the characteristics obtained under the static tensile tests. The hybrid nanocomposites are the exception since tensile strength in the case of cyclic tests is greater than the tensile strength of corresponding nanocomposites with one type of filler (Figure 12). As shown in Figure 13, the value of modulus at 200% strain during cyclic tests, is the highest for hybrid nanocomposite containing 6 wt.% of CNTs and HNTs (over 5.5 MPa). EVA/4 wt.% CNTs + HNTs also achieves higher modulus at 200% strain value than the corresponding nanocomposites with CNTs or HNTs. These differences may result from the orientation of nanotubes along with the tensile direction. After the removal of the applied force, nanotubes did not return to the original position. Thus, the addition of HNTs enables the material to reach higher strains by its plasticizing effect, while CNTs align upon cycling straining and give a higher gain in terms of tensile strength. This reveals an intriguing strengthening mechanism, which occurs because of synergism of both nanotube fillers and which can be used in the development of nanocomposites with behavior allowing to maintain higher cyclic strains and loads. It can be seen that nanocomposites containing CNTs show higher values of the permanent set than the neat EVA copolymer (Figure 14). The EVA/8wt.% CNT nanocomposite showed the highest value of permanent set (PS(200%)) of over 85%, which was about two times higher than that of neat EVA copolymer. As mentioned earlier, CNTs increase the stiffness of the composite, which can contribute to higher residual strain values. HNTs behave the opposite way. As can be seen in Figure 12b, EVA/HNTs nanocomposites have slightly better recovery properties, than the neat EVA copolymer.

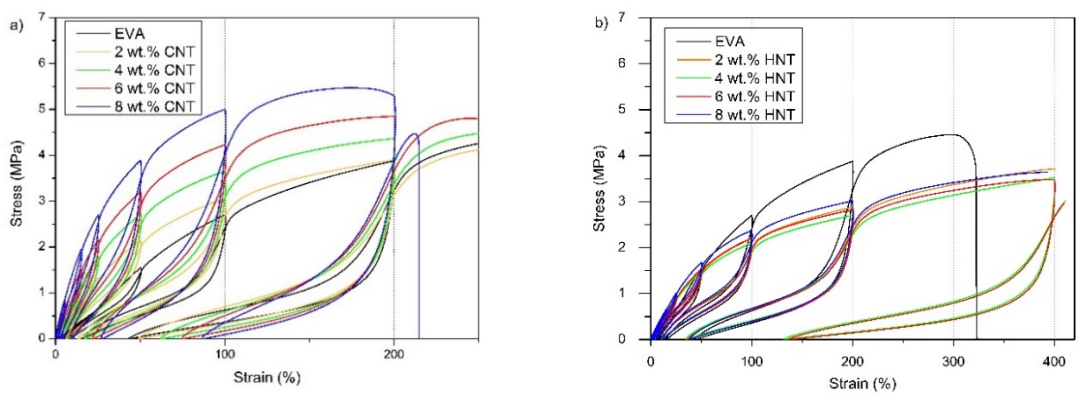

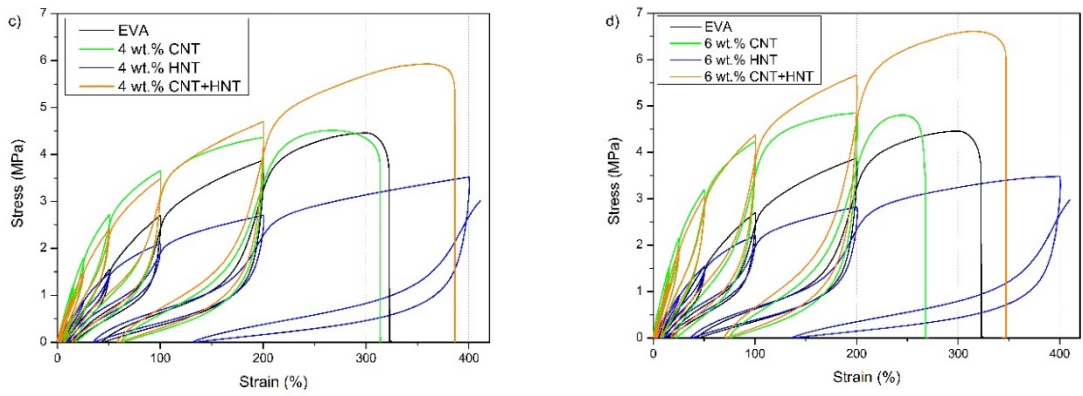

**Figure 12.** The stress–strain curves under cyclic loading for EVA/MWCNT (**a**), EVA/HNT (**b**), EVA/CNT + HNT (**c–d**) nanocomposites, and the neat EVA copolymer under. Results corresponding to 5%, 15%, 25%, 50%, 100%, and 200% maximum tensile strains.

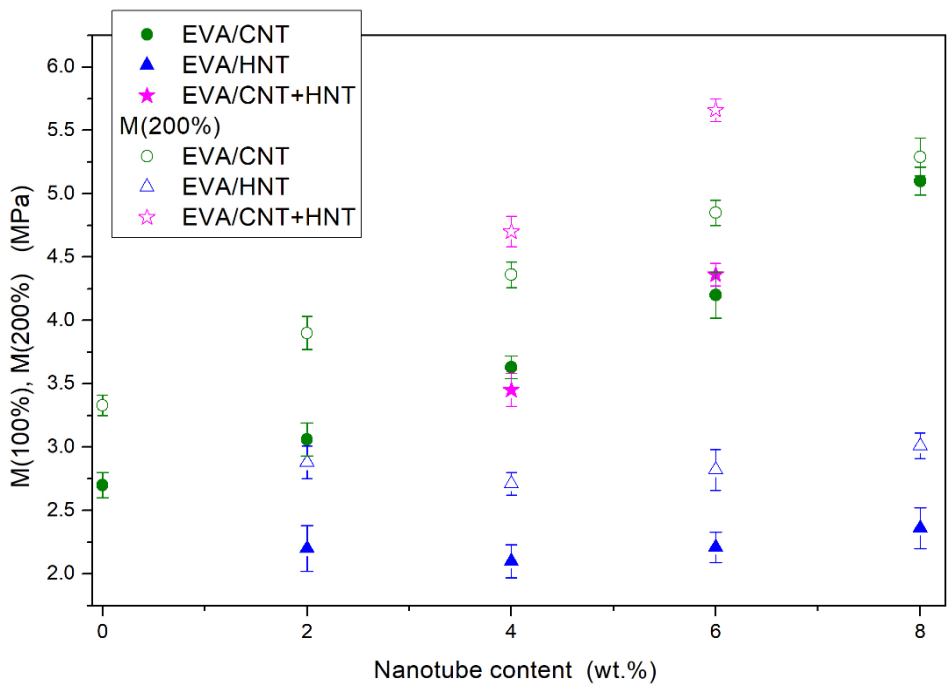

**Figure 13.** Modulus at 100 and 200% of elongation for the neat EVA and EVA-based nanocomposites after maximum strain attained in the cycle.

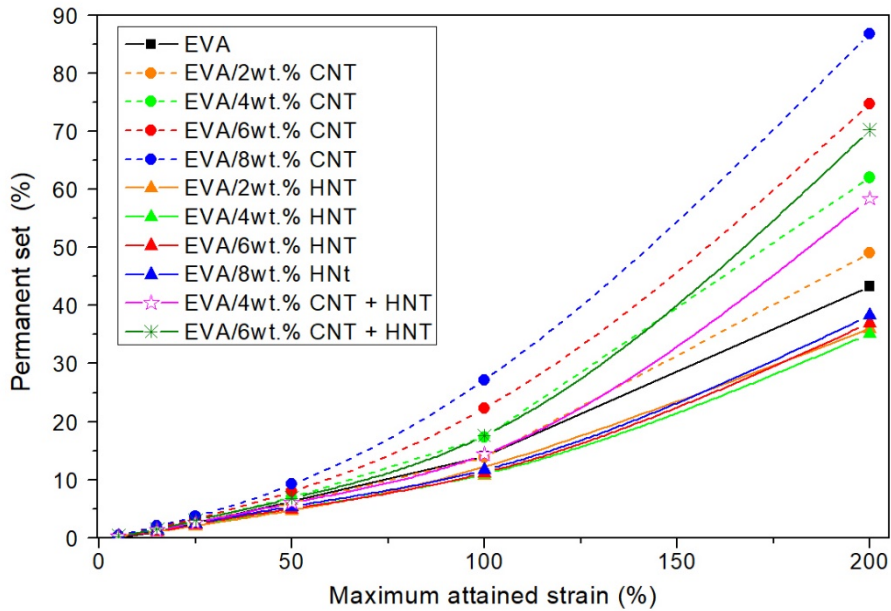

**Figure 14.** Permanent set values of the neat EVA, and EVA-based nanocomposites after maximum strain attained in a cycle.

*3.5. Electrical Conductivity of EVA Nanocomposites*

MWCNTs possess very high intrinsic electrical conductivity. Even their small content can significantly improve the electrical conductivity of insulating polymers. Figure 15 displays the electrical properties of EVA-based nanocomposites. It can be seen that the electrical conductivity increased steadily in EVA/CNTs nanocomposites, as the CNTs concentrations increased. The highest electrical conductivity value of $5.2 \cdot 10^{-7}$ S/m was achieved for nanocomposite with 8 wt.% of CNTs. It is almost 6 orders of magnitude higher than for neat EVA copolymer. Obtained materials did not show a sharp increase in electrical conductivity along with the increase in MWCNTs content, in contrast to the results from previous work in which the matrix was low-density polyethylene (LDPE) [78]. In LDPE/MWCNTs nanocomposites the percolation threshold was observed at the loading of 1.5 wt.% of the same MWCNTs [78]. It is known that relatively uniform dispersion of CNTs can be achieved in polar polymers such as polyamide, polycarbonate because of the strong interaction between the polar moiety of the polymer chains and the surface of the CNTs [79]. EVA copolymer is polar in comparison to the LDPE and its polarity is dependent on the mass content of VA in the copolymer. Because of the interactions between polar groups of EVA copolymer and nanotubes, the carbon nanotubes may be more concentrated in VA polar domains. Therefore, in EVA nanocomposites with the highest concentration (8 wt.%) of nanotubes, regions/agglomerates with a higher content of entangled of nanotubes are visible (Figure 2).

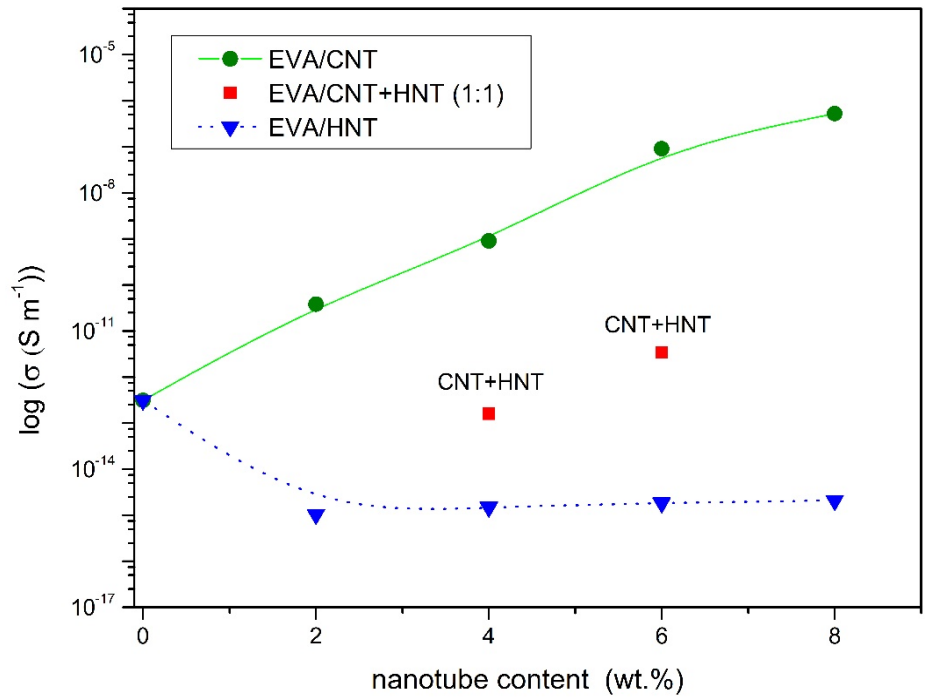

**Figure 15.** Electrical conductivity of EVA/CNT, EVA/HNT, and EVA/CNT + HNT nanocomposites.

Because halloysite nanotubes are not conductive fillers, the EVA/HNTs nanocomposites exhibit lower electrical conductivity than a neat EVA copolymer. As shown in Figure 15, the conductivity decreased by two orders of magnitude. Moreover, the insulating properties of HNTs cause no significant improvement in the conductivity of hybrid nanocomposites. Hybrid material at the concentration of 4 wt.% exhibits an even slightly lower value of conductivity if compared to neat EVA. This can be explained by the fact that HNTs located between CNTs, impedes the formation of conductive pathways.

## 4. Conclusions

Nanocomposites based on EVA copolymer containing MWCNTs, HNTs, or both types of nanotubes were prepared by melt blending. One found that relatively good nanotubes distribution in polymer matrix were obtained. In the case of EVA/CNT nanocomposites at high load (6–8 wt.%), the MWCNTs form the highly exfoliated network structure in EVA matrix, in which individual nanotubes and entangled CNTs in agglomerates are present. DSC studies showed that the addition of nanofillers caused no significant effect on the melting temperature and the degree of crystallinity. However, glass transition temperature slightly increased. At high temperatures ($T_{50\%}$), nanocomposites showed better thermo-oxidative stability than the neat EVA copolymer. A slight improvement in thermal stability was also noted. Moreover, the addition of CNTs has significantly improved the mechanical properties of EVA copolymer. Nanocomposites were stiffer and their tensile strength increased by about 40%. In turn, HNTs give an opposite effect in terms of composite stiffness and tensile strength. They decrease the strength of EVA. However, the strain to break increases by over 70% when HNTs are added. Furthermore, the cyclic tensile tests demonstrated that nanocomposites with HNTs have slightly better recovery properties, than pure EVA. Interestingly, in cyclic tensile tests significant improvement of tensile strength for hybrid nanocomposites was visible. Moreover, the extraordinary strengthening caused by synergism of the used nanotube fillers did not diminish the strain rates achieved by the hybrids. Nanocomposites with CNTs were found to be electrically conducting. For nanocomposites containing 8 wt.% of CNTs, an increase in electrical conductivity for about six orders of magnitude in comparison to the neat copolymer was observed.

**Supplementary Materials:** The following are available online at www.mdpi.com/xxx/s1. Figure S1: TEM images of EVA/6 wt.% CNT + HNT nanocomposite at different magnifications. Figure S2: TEM images of EVA/6 wt.% CNTs nanocomposite at different magnifications.

**Author Contributions:** A.Z. prepared the literature review, analyzed the results, wrote the original draft of the manuscript, and performed physical properties measurements; A.S. planned the experiment, supervised the discussion, and reviewed the manuscript; P.F. prepared the samples, analyzed the mechanical results, and reviewed the manuscript; A.K. performed the SEM experiments; I.J. participated in TEM analysis and supervised the discussion of the results; S.P. analyzed the thermal properties of the samples and participated in editing and revision of the manuscript. All authors have read and agreed to the published version of the manuscript.

**Funding:** This research received no external funding.

**Acknowledgments:** The authors wish to thank Walid Baazis from IPCMS (UMR 7504 CNRS-UDS, Strasbourg) for the TEM investigations. The authors also would like to express their appreciation to PRACHT GROUP for the long-term loan of Arburg Allrounder 270 S 350–100 injection molding machine.

**Conflicts of Interest:** The authors declare no conflict of interest.